\documentstyle
[preprint,tighten,floats,prl,aps]{revtex}

\input
epsf
\begin{document}

\draft
\title{Prospects for SIMPLE 2000: A large-mass, low-background
Superheated Droplet Detector for WIMP searches}
\author{J.I. Collar$^{a,b,*}$, J. Puibasset$^{a}$, T.A. Girard$^{c}$, D.
Limagne$^{a}$, H.S. Miley$^{d}$ and G. Waysand$^{a,e}$}
\address{
$^{a}$Groupe de Physique des Solides (UMR CNRS 75-88), Universit\'es Paris
7 \& 6,75251 Paris Cedex 05, France\\
$^{b}$CERN, EP Division, CH-1211
Geneve 23, Switzerland\\
$^{c}$Centro de F\'\i sica Nuclear, Universidade
de Lisboa, 1649-003 Lisbon,
Portugal\\
$^{d}$Pacific Northwest National
Laboratory, Richland, WA 99352,
USA\\
$^{e}$Laboratoire Souterrain $\grave{a}$ Bas Bruit de Rustrel-Pays d'Apt,
Rustrel 84400, France\\
}
\maketitle
\begin{abstract}
\widetext
SIMPLE 2000 ({\underline S}uperheated {\underline I}nstrument 
for {\underline
M}assive {\underline P}artic{\underline {LE}} searches) 
will consist of an
array of eight to sixteen large active mass ($\sim\!15$ g) Superheated
Droplet Detectors(SDDs) to be installed in the new underground
laboratory of Rustrel-Pays d'Apt. Several factors make of SDDs 
an attractive approach for the detection of Weakly Interacting
Massive Particles (WIMPs), namely their
intrinsic insensitivity to minimum 
ionizing particles,
high fluorine content, low cost and operation near ambient 
pressure and temperature. We comment here on the 
fabrication, calibration
and already-competitive first limits from SIMPLE 
prototype SDDs, as well as
on the expected immediate increase in sensitivity of the program,
which aims at an exposure of $>$25 kg-day during the
year 2000. The ability of modest-mass fluorine-rich detectors 
to explore regions of neutralino parameter space 
beyond the reach of the most ambitious cryogenic projects is 
pointed out.
\end{abstract}
\pacs{PACS number(s):
95.35.+d, 29.40.-n,
05.70.Fh\\
$^{*}$~Corresponding author. E-mail:
Juan.Collar@cern.ch}
\narrowtext
Apfel extended
the Bubble Chamber concept
with the invention of
Superheated Droplet Detectors \cite{apfel1} (SDDs,
also known as Bubble Detectors), a dispersion of small drops (radius
$\sim10~\mu m$)
of superheated liquid in a gel or viscoelastic matrix.
The SDD matrix
isolates the fragile metastable system from vibrations and
specially
from convection
currents (absent in gels), while the smooth
liquid-liquid interfaces
impede the continuous triggering on surface
impurities
that occurs in the walls and gaskets of
even the cleanest
bubble chambers. A physically-sound metaphor for the
gained SDD stability
over bubble chambers is the improvement
that was brought on by dynamite
over nitroglycerine: in SDDs, the lifetime
of the superheated system
is
extended to the point that new practical applications such as
personnel
and area neutron dosimetry become possible.

In the moderately superheated
industrial refrigerants used in SDDs,
bubbles can  be produced only by
particles having elevated
stopping powers ($dE/dx\gtrsim 200 ~keV/\mu m$)
as is the case for low-energy
nuclear recoils. This behavior is described
by Seitz's classical
``thermal spike'' model \cite{seitz}: for the
transition to occur, a vapor
nucleus or ``protobubble'' of radius $>
r_{c}$ must be created, while only the energy
deposited along a distance
comparable to this critical radius $r_{c}$
is available for its formation.
Hence, a double
threshold is imposed: the requirement that the deposited
energy
$E$
be larger than the thermodynamical
work of formation of the
critical nucleus, $E_{c}$, and that this
energy be lost by
the particle
over a distance $\textrm{O}\!\left(r_{c}\right)$,
i.e., a minimum stopping
power. Protobubbles formed by energy depositions not meeting both
demands
simply shrink back to zero; otherwise, the transition is
irreversible and
the whole droplet vaporizes. Formally expressed,
these two conditions
become \cite{apfel2,peyrou}:
\begin{eqnarray}
E>E_{c}=4\pi r^{2}_{c}
\gamma/3\epsilon \nonumber\\
dE/dx>E_{c}/a r_{c},
\end{eqnarray}
where
$r_{c}\!=\!2~\gamma/\Delta P$,
$\gamma\!\left(T\right)$ is the surface
tension,
$\Delta P\!=\!P_{V}\!-\!P$, $P_{V}\!\left(T\right) $
is the vapor
pressure of the liquid (generally an industrial refrigerant), $P$ and $T$ are the operating pressure and
temperature, $\epsilon$ varies in the range
$[0.02,0.06]$ for different
liquids \cite{peyrou,apfelroy}, and $a\!\left(T\right)\!\sim \!
\textrm{O}\!\left(1\right)$ \cite{harper}. The parameter $\epsilon$ is
of
particular importance in the calculation of the minimum recoil energy
that
needs to be transferred in a neutron (or WIMP) collision for
a bubble to
form, given that in most cases these recoils have
a sufficiently high
$dE/dx$ to pass the second requirement. The
physical meaning of $\epsilon$
is often misinterpreted
by modern authors  as a somewhat undefined
fraction of deposited energy available for
protobubble formation. The
reality (well-known during the bubble
chamber era) is different: making $\epsilon\!=\!1$,
$E_{c}$
as calculated
in \frenchspacing{Eq. (1)} becomes equal to
the minimum (Gibbs) bubble expansion 
work, a gross
but convenient underestimate of
the actual work of bubble formation.
Examining the process in detail,
$E_{c}$  can be recalculated as the much
larger sum of the reversible
works of bubble surface formation,
evaporation of the liquid and expansion against $P$ \cite{peyrou,harper}:
\begin{equation}
E'_{c}=4\pi r^{2}_{c} \left(\gamma - T\frac{\partial
\gamma}{\partial T}
\right)+\frac{4}{3}\pi r^{3}_{c} \rho_{v}
\frac{h_{fg}}{M}+\frac{4}{3}\pi
r^{3}_{c} P,
\end{equation}
where
$\rho_{v}$ is the saturated vapor density, $h_{fg}$ is the
latent heat of
vaporization per mole and $M$ the molecular mass. By
making
\begin{equation}
\epsilon=4\pi r^{2}_{c} \gamma/3E'_{c},
\end{equation}
\frenchspacing{Eq. (1)} acquires a compact form but leads
to the
mentioned conceptual misunderstanding.
It must be emphasized that
$\epsilon$ as correctly defined in \frenchspacing{Eq. (3)} is not a 
free parameter: it can be
accurately
calculated for each refrigerant and operating conditions
(\frenchspacing{Fig. 1}) and compared with experimental measurements
(see below).

The second condition in \frenchspacing{Eq. (1)} can
be exploited
to avoid the
background from ubiquitous minimum-ionizing radiation that
plagues experiments aiming to detect
WIMP-induced nuclear recoils (WIMPs
are one of the best
candidates for
the galactic dark
matter
\cite{wimps}). SDDs of active mass O(1)kg can in principle
considerably
extend the present experimental
sensitivity  \cite{myprd} well into the
region where new supersymmetric
particles are expected. The low
interaction rate expected
from these particles
($<\!\!10$ recoils/kg
target mass/day) and the modest active
mass of commercially available SDDs
($\sim0.03$ g refrigerant/dosimeter), together with a desire to
control
and understand the fabrication process,
lead us to develop in
collaboration with COMEX-PRO \cite{comex} a large-volume ($
80~l$)
pressure reactor dedicated to
SDD production. Able to withstand 60 atm,
it
houses a variable-speed magnetic stirrer, heating and
cooling elements
and micropumps for the addition of
catalysts whenever chemical
cross-linking of the gel is required.
We have nevertheless favored
thermally-reversible food gels
such as agarose, gelatine,
$\kappa$-carrageenan, etc.,
due to safety
concerns in the handling of
large volumes of synthetic monomers.
The fabrication of $1~l$ SDD
modules
containing up to 3\% in superheated liquid starts with the
preparation of
a suitable gel matrix;
the constituent materials must be carefully
selected and processed in order to avoid alpha emitters,
the only
internal radioactive contaminants of concern
\cite{myprd}. A still with
all contact parts made of quartz and teflon is used to produce
high-purity bidistilled water. Unfortunately, a very precise density
matching between the matrix and refrigerant is needed to obtain a
uniform
droplet dispersion, making
water-based gels inadequate unless large
fractions of inorganic
salts are added, which can unbalance the chemistry
of the composite
and contribute an undesirable concentration of
alpha-emitters
\cite{zacek2}. We find that glycerol is for this and
other
reasons an additive of choice.
The gelating agent, polymer additives
and glycerol are purified using
a pre-eluted ion-exchanging resin
specifically targeted at
actinide removal. All components are forced
through
0.2 $\mu m$ filters to remove motes that
might act as nucleation
centers. The resulting mixture is outgassed
and maintained in the
reactor
above its gelation temperature. The refrigerant is
single-distilled
prior to its incorporation to this solution, which is 
done at a pressure
well above $P_{V}\!\left(T\right)$ to avoid boiling during the vigorous
stirring that follows (\frenchspacing{Fig. 2}). After a uniform dispersion
of droplets is
obtained, cooling, setting and step-wise adiabatic
decompression produce
the delicate entanglement of superheated liquid
and thermally-reversible gel that makes up the SDD.
Numerous practical
precautions, to be described elsewhere, go into producing
stable modules;
for instance, the step-wise decompression procedure
used is identical to
that employed by scuba-divers returning
to the surface, in order to minimize the cavitation of
dissolved gas
bubbles which in SDDs can
act as inhomogeneous nucleation centers.
The detectors are refrigerated
and pressurized under 4 atm
during storage and transportation, to inhibit
their response to
environmental
neutrons.

While SDDs can bypass the
mentioned problems
associated to a former \cite{zacek} bubble chamber WIMP
search proposal,
they are not devoid of their own idiosyncrasies. During
the R\&D
leading to the first SIMPLE modules, we have been able to
identify
and solve some of the SDD particularities that can interfere with
a successful WIMP search. For instance, the appearance of fractures in
the gel
and depletion of active mass over long exposures via permeation
processes, or the formation of clathrate-hydrates at the droplet
boundaries during fabrication or recompression (these are crystalline
structures able to destroy the metastability of the droplets). These
detector improvements have been treated at some
length elsewhere
\cite{ourprletc}
and are summarized here in table 1. The objective was to
keep the
detector components down to a minimum for reasons of radiopurity,
safety and cost, while solving the condensed-matter issues as they
appeared. Of special mention are the gains in detector stability brought
by the transition from R-12
($CCl_{2}F_{2}$) to R-115 ($C_{2}ClF_{5}$)
(see table 1). Present
SIMPLE modules contain $\sim 1000$ times the active
mass of available
commercial SDDs (limited only by the size of the
pressure reactor)
and can be operated continuously for up to $\sim40$ d. 
Even though a further extension of this
shelf life is intended, the design of recompression 
chambers that would allow for 
an indefinite exposure is under study. The cost of the SDD matrix
has been kept down, allowing for a much
larger future
design.

\begin{table}
\caption{Characteristics of present SIMPLE SDDs.}
\begin{tabular}{|c|l|l|}
\hline
 & & \\
\bf{component} & \bf{ features} & \bf{                   goal}\\
 & & \\
\hline
 & & \\
water & $\bullet$ immiscible with most H-free 
refrigerants & \textsc{chemical
compatibility} \\
 &  $\bullet$ bi-distilled (quartz and 
teflon still) & \textsc{radiopurity} \\
 & & \\
\hline
 & & \\
 &  $\bullet$ density matching between sol \& freon & \textsc{SDD 
homogeneity} \\
 &  $\bullet$ viscosity enhancer & \textsc{fracture control (diffusion
$\downarrow$)}\\
 &   $\bullet$ solvent behavior similar to water & \textsc{chemical
compatibility} \\
glycerin &  $\bullet$ excellent wetting of glass 
surfaces & \textsc{absence of
nucleations on walls} \\
 &  $\bullet$ low U+Th content and easy to 
purify & \textsc{radiopurity} \\
 &  $\bullet$ does not crystallize at low 
T & \textsc{lack of inhomogeneous
nucleations} \\
 & $\bullet$ germicide & \textsc{no need to add preservatives} \\
 & $\bullet$ strengthens gelatine gels & \textsc{structural stability} \\
 & & \\
\hline
 & & \\
 & $\bullet$ viscosity enhancer & \textsc{fracture control} \\
 & $\bullet$ surfactant & \textsc{SDD homogeneity} \\
 & $\bullet$ salting-out agent at low \% & \textsc{fracture control
(solubility $\downarrow$}) \\
PVP &  $\bullet$ kinetic inhibitor at 
low \% & \textsc{absence of clathrate-hydrates} \\
 &  $\bullet$ compatible with gelatine 
gels & \textsc{chemical compatibility} \\
 &  $\bullet$ chelating agent & \textsc{stops $\alpha$-emitter
migration to} \\
 &   & \textsc{  droplet boundaries} \\
 & & \\
\hline
 & & \\
 &  $\bullet$ forms compliant
gel in $T$ range & \textsc{structural stability} \\
gelatine &  $\bullet$ non-toxic & \textsc{safety} \\
 &  $\bullet$ low U+Th (depends on organ
origin) & \textsc{radiopurity} \\
 & & \\
\hline
 & & \\
 &  $\bullet$ 62\% F (twice as much as R-12) & \textsc{higher WIMP
sensitivity} \\
 &  $\bullet$ non-toxic, non-flammable & \textsc{safety} \\
 &  $\bullet$ much lower solubility than R-12 & \textsc{control of
fractures and } \\
R-115 &   & \textsc{  active-mass losses} \\
($C_{2}ClF_{5}$) &  $\bullet$ much larger
molecule than R-12 & \textsc{clathrates impossible,
diffusion $\downarrow$} \\
 &  $\bullet$ higher $P_{V}$ than R-12 & \\
 &  $\Rightarrow$ allows operation at 
lower $T$ & \textsc{fracture control,
diffusion $\downarrow$} \\
 & & \\
\hline
 & & \\
- & $\bullet$ operation at 2 atm  & \textsc{fracture control} \\
 &  & \textsc{ + keeps radon out} \\
\end{tabular}
\end{table}
Prototype modules have been tested in an
underground
gallery 40 km south of Paris. The 27 m rock
overburden and
$\sim\! 30$ cm paraffin shielding reduce
the flux of cosmic and
muon-induced fast neutrons,
the main source of
nucleations above ground.
Inside the shielding, a water$+$glycol thermally-regulated
bath maintains
$T$ constant to within $0.1^{\circ}$C.
The characteristic violent sound
emission accompanying vaporization
in superheated liquids
\cite{sound1,sound2}
is picked-up by a small piezoelectric transducer in
the
interior of the module, amplified
and saved in a storage
oscilloscope. Special precautions are
taken against acoustic and seismic
noise.
\frenchspacing{Fig. 3} displays the decrease
in spontaneous bubble
nucleation rate
brought by progressive purification of
the modules, as
measured in this site. The ability to
rapidly modify the fabrication
process and the choice
of components has been critical in achieving
this.

The response of smaller SDDs to various
neutron fields has been
extensively studied \cite{harper,apfel3,derico}
and found to match theoretical
expectations. However, large-size, opaque SDDs
require independent
calibration:
acoustic detection of the explosion of the smallest or most
distant
droplets is not {\it a priori} guaranteed. The
energy released
as sound varies as
$(P_{V}-P)^{3/2}$ \cite{sound2}, making these
additional
characterizations even more imperative
for SDDs operated under
moderate $P$. Two separate
types of calibration have
been performed to
determine the target mass  effectively monitored
in SIMPLE modules and to
check the calculation of the $T,P$-dependent
threshold energy $E_{thr}$
above which WIMP recoils can induce
nucleations (defined as the lowest
energy meeting both
conditions in \frenchspacing{Eq. (1)}
\cite{myprd,nagdy}).
First, a liquid $^{241}$Am source
(an alpha-emitter)
is uniformly diluted into the matrix prior to gel setting.
Following
Eq. (1), the 5.5 MeV alphas and 91 keV recoiling $^{237}$Np daughters
cannot induce nucleations at temperatures below $T_{\alpha}$ and
$T_{\alpha r}$, respectively \cite{myprd}.
The expression  $a=4.3
\left(\rho_{v}/\rho_{l}\right)^{1/3}$ \cite{harper},
where 
$\rho_{l}(T)$ is the saturated liquid 
density of the refrigerant,
correctly predicts
the observed $T_{\alpha}$ for both R-12 and
R-115 at $P\!=$1 and 2 atm. Under similar conditions,
the theoretical value of
$\epsilon$ for
these liquids (\frenchspacing{Fig. 1})
accurately predicts the experimental $T_{\alpha r}$
(\frenchspacing{Fig. 4}), which 
directly depends on it.
Calibrations like those in \frenchspacing{Fig. 4} allow to corroborate
the calculation of $E_{c}$ for each refrigerant: a good knowledge of 
$E_{c}$ is crucial to
determine the expected
SDD sensitivity to WIMP recoils at
different
operating $P,T$.

Prior to extensive component purification,
the spectrum
in non-calibration runs (\frenchspacing{Fig. 3}, histogram)
had a close
resemblance to that produced by $^{241}$Am
spiking;
the initial presence
of a small ($\sim\! 10^{-4}$ pCi/g) $^{228}$Th contamination,
compatible
with the observed rate, was confirmed
via low-level alpha spectroscopy.
Three regimes of background
dominance are therefore delimited by vertical
lines in \frenchspacing{Fig.
3}: the sudden rise
at $T\!\sim\!15^{\circ}$C originates in 
high-$dE/dx$ Auger electron cascades
following interactions of environmental gammas with Cl atoms in the
refrigerant \cite{peyrou,hahn,tenner}. 
The calculated $E_{c}$ for R-115 at
$T\!=\!15.5^{\circ}$C and
$\!P\!=$2 atm is 2.9 keV, coincidental with the
binding energy of K-shell
electrons in Cl, 2.8 keV (i.e., the maximum $E$
deposited
via this mechanism). Thus, the onset of gamma sensitivity
provides an
additional check of the threshold in the few keV
region.

Regrettably, alpha calibrations cannot be used for a rigorous
determination of the overall sound detection efficiency because a large
fraction of the added emitters drifts to gel-droplet boundaries
during fabrication, an effect explained by the polarity of actinide
complex ions
\cite{wang}. We observe that this effect 
is seemingly dependent on matrix 
composition (the migration can be 
controlled to some extent 
by the addition of chelating polymers such as PVP, which can link 
to the actinides while becoming themselves immobilized 
by entanglement in 
the gel structure).
While this migration does
not alter the expected value of $T_{\alpha}$ nor $T_{\alpha r}$, it
enhances the overall
nucleation efficiency in a somewhat unpredictable
manner \cite{wang}. To
complete our understanding of the detector efficiency, SIMPLE modules
have been exposed to a
well-characterized $^{252}$Cf neutron
source at the TIS/RP calibration
facility (CERN). The resulting
spectrum of neutron-induced fluorine
recoils
(\frenchspacing{Fig. 5}, insert) mimics
a typically expected one
from WIMP interactions. A complete MCNP4a
\cite{mcnp} simulation of the
calibration setup takes into account
the small contribution from albedo and
thermal neutrons.
The expected nucleation rate as a function of $T$ is
calculated as in
\cite{myprd,apfel3}: cross sections for the elastic,
inelastic,
(n,$\alpha$) and (n,p) channels of the refrigerant constituents
are extracted from ENDFB-VI libraries. Look-up
tables of the distribution
of deposited energies as a function of
neutron energy are built from the
SPECTER code \cite{specter} and
stopping powers of the recoiling species are
taken from SRIM98
\cite{trim}.
Since $T$ was continuously ramped up during
the irradiations at
a relatively fast 1.1$^{\circ}$C/hr, a small
correction to it ($<\!1^{\circ}$C) is
numerically computed and applied to
account for
the slow thermalization of the module.
Depending on $T$, the value 
of $E_{thr}$ for fluorine elastic recoils (the dominant nucleation 
mechanism in R-115) is set by either condition in \frenchspacing{Eq. 
(1)}, the other being always fulfilled for $E>E_{thr}$
\cite{myprd,nagdy}. 
The handover from the second to the
first condition at $T$ above $\sim 5.5^{\circ}$C ($\sim 2.5^{\circ}$C) for 
$P\!=$2 atm ($P\!=$1 atm) 
is clearly observed in the data 
as two different regimes of nucleation rate (\frenchspacing{Fig. 5}). 
A larger-than-expected response, already noticed 
in R-12 \cite{harper}, is evident at low $T$: the calculated $E_{thr}$ 
there is too conservative (too high). This behavior appears well 
below the normal regime of SDD operation (which is at $T$ high 
enough to have $E_{thr}\!=\!E_{c}$) and therefore does not interfere with 
neutron or WIMP detection. However, it is interesting in that it 
points at a higher than normal bubble nucleation 
efficiency from heavy particles, 
as discussed in early bubble chamber work \cite{tenner}.
It is precisely at low $T$ that
the spontaneous nucleation
rate in low-background conditions is
the smallest. Therefore this effect,
which could greatly improve SDD
WIMP limits, merits further attention: calibrations
using filtered neutron beams of
energies $2\pm 0.8$, $24.3\pm 2$, $55\pm 2$ and $144\pm
24$ keV available 
at the research reactor of the Nuclear and Technology
Institute (Sacavem, Portugal) are
planned as part of the
SIMPLE 2000 effort.
A best-fit to the overall normalization of the Monte Carlo over the 
full data set (\frenchspacing{Fig. 5}) enables 
us to determine the refrigerant mass 
monitored with the present sound acquisition chain as 
$34\pm 2\%$ ($74\pm 4\%$) of the total at $P\!=$2 atm ($P\!=$1 atm), 
a decisive datum to obtain dark matter limits.

The installation 500 m underground of modules identical in
composition,
preparation and sound detection system to those utilized in
$^{252}$Cf
calibrations started in July 1999 (\frenchspacing{Fig. 6}).
A decommissioned nuclear missile launching
control center has been converted into an
underground laboratory
\cite{wwwrustrel}, facilitating this and other
initiatives.
The characteristics of this site (microphonic silence, unique
electromagnetic shielding of the halls \cite{wwwrustrel}) make it
specially
adequate for rare-event searches. Modules are placed inside a
thermally-regulated water bath, surrounded by three layers of sound and
thermal insulation. A 700 l water pool acting as neutron moderator,
resting on a dual vibration absorber, completes the shielding. Events in
the modules and in external acoustic and seismic monitors are time-tagged,
allowing to filter-out the small percentage ($\sim\!15$\%) of signals
correlated
to human activity in the immediate vicinity of the experiment.
The signal waveforms are digitally stored, but no event rejection based
on pulse-shape considerations \cite{zacek2} is performed at this stage,
avoiding the criticisms \cite{gerbier} associated to some WIMP searches
in which large cuts to the data are made. The raw counting rate from the
first SIMPLE module operated in these conditions appears in
\frenchspacing{Fig. 7}.  Accounting for sound
detection efficiency and a
62\% fluorine mass fraction in R-115,
limits can be extracted
on the
spin-dependent WIMP-proton cross section $\sigma_{Wp}$
(\frenchspacing{Fig. 8}).
The cosmological parameters and method
described in \cite{smith} are
used in the calculation of WIMP elastic
scattering rates, which are then
compared to the observed uncut nucleation
rate at $10^{\circ}$C ($14^{\circ}$C
for small WIMP masses). The expected
nucleation rate
at $T$ (i.e., integrated for recoil energies above
$E_{thr}(T)$) from a candidate at the
edge of the sensitivity of the
leading DAMA experiment \cite{review}
($\sim\!1.5\cdot10^{4}$ kg-day of
NaI)
is offered as a reference in \frenchspacing{Fig. 7}: evidently, with
the same level of background but significantly 
smaller statistical error bars,
this candidate could have
been marginally excluded.
Present SIMPLE limits are
impaired
by the large statistical uncertainty associated to the short
exposure
accumulated so far, and not yet by background 
rate. A considerable improvement
is expected after the
ongoing expansion of the bath to accommodate up to
16 modules (\frenchspacing{Fig. 8}). SIMPLE 2000
aims at an exposure of $\sim 25$ kg-day in the
next few
months, by
replacing  the detectors (in batches of eight) every four to six
weeks,
repeating this cycle several times.
A weak Am/Be neutron source will be
used at the end
of each run to assess the sound detection efficiency for
each module
{\it in situ}. In parallel to this, plastic module caps are
being replaced by
a sturdier design: runs using refrigerant-free modules
show that
{\it a majority} of the prototype events arose from pressure
microleaks, correlated
to the sense of $T$ ramping,
able to stimulate
the piezoelectric sensor (\frenchspacing{Figs. 7 and 9}). In principle,
if this
source of background is controlled, the maximum sensitivity
of SIMPLE 2000
can start to probe the spin-dependent neutralino parameter space
(\frenchspacing{Fig.
8}). It must also
be kept in mind that a $T-$independent, flat background
implies a
null WIMP signal, albeit this eventual approach to data analysis
can only be
exploited after a sizable reduction in statistical uncertainty
is achieved.

The importance of the spin-dependent WIMP interaction channel,
for which fluorine-rich detectors are 
by far the optimal target \cite{john}, has
been recently
underlined by its relative insensitivity to CP-violation
parameter
values, which may otherwise severely reduce coherent (i.e., 
spin-independent) 
interaction
rates \cite{paolo}. To further stress the significance of this channel,
we illustrate in \frenchspacing{Fig.
10} a not-so-obvious
complementarity
of spin-dependent and spin-independent
searches in exploring the neutralino
phase space.
The top-left frame displays
points generated
with the help of the NEUTDRIVER code \cite{wimps}, 
each representing a
possible combination of MSSM parameters. The parameter space sampled
is the same as in \cite{bottino} (namely, 10 GeV$\leq\! 
M_{2}\!\leq$ 10 TeV, 10 GeV$\leq \mid\!\mu\!\mid \leq$ 10 TeV, 1.1 $\leq$ 
tan($\beta$)$\leq$ 50, 60 GeV$\leq\!
m_{A}\!\leq$ 1 TeV, 100 GeV$\leq 
m_{0}\leq$ 1 TeV) and special precautions are taken to do so as 
homogeneously as possible, within the limitations imposed by 
computing time. Only a weak
correlation between
the values of $\sigma_{Wp}$ (i.e., the spin-dependent
coupling strength, corrected for local halo density)
and $\sigma_{Wn}$ (spin-independent) is observed in the plot (note 
the scale).
As a result of this, a compact
($\sim 1$ kg active mass), low-background SDD
starting to
probe the neutralino $\sigma_{Wp}$ (top-right frame) generates a rather
homogeneous ``cleaning'' of the MSSM models 
in a $\sigma_{Wn}$
exclusion plot (bottom-left frame,
without the constrains imposed
by the limits in top-right; bottom-right with them), 
down to values of $\sigma_{Wn}$
far beyond 
the reach of the most ambitious planned cryogenic WIMP searches. This
can be counter-intuitive for
the hardcore experimentalist,
which might otherwise naively expect the SDD to take
only a small bite
off the top of the $\sigma_{Wn}$ cloud in the
lower-left frame (in
other words, might expect much more of a correlation
between
$\sigma_{Wp}$ and $\sigma_{Wn}$).
Needless to say, the converse
can be stated of the way that 
cutting-edge spin-independent searches will 
deplete the
$\sigma_{Wp}$
cloud and hence the complementarity of both approaches. 
In conclusion, 
if an exhaustive test of the neutralino-as-cold-dark-matter hypothesis
is ever to be achieved, the development of fluorine-rich 
detectors cannot be neglected: in this respect SDDs represent an 
ideal opportunity.

We thank the
Communaut\'e des Communes du Pays d'Apt
and French Department of Defense
for supporting the
conversion of the underground site. G. Jungman graciously provided 
the NEUTDRIVER code and assistance in implementing it. 
Our gratitude also
goes to M. Auguste,
J. Bourges, G. Boyer, R. Brodzinski, A. Cavaillou,
COMEX-PRO,
M. El-Majd, M. Embid, L. Ibtiouene,
IMEC, J. Matricon, M. Minowa, Y.H. Mori, T. Otto,
G. Roubaud, M. Same and C.W.
Thomas.
J.I.C. was supported by the EU TMR
programme.

\newpage
\begin{figure}[tbp]
\epsfxsize = \hsize 
\epsfbox {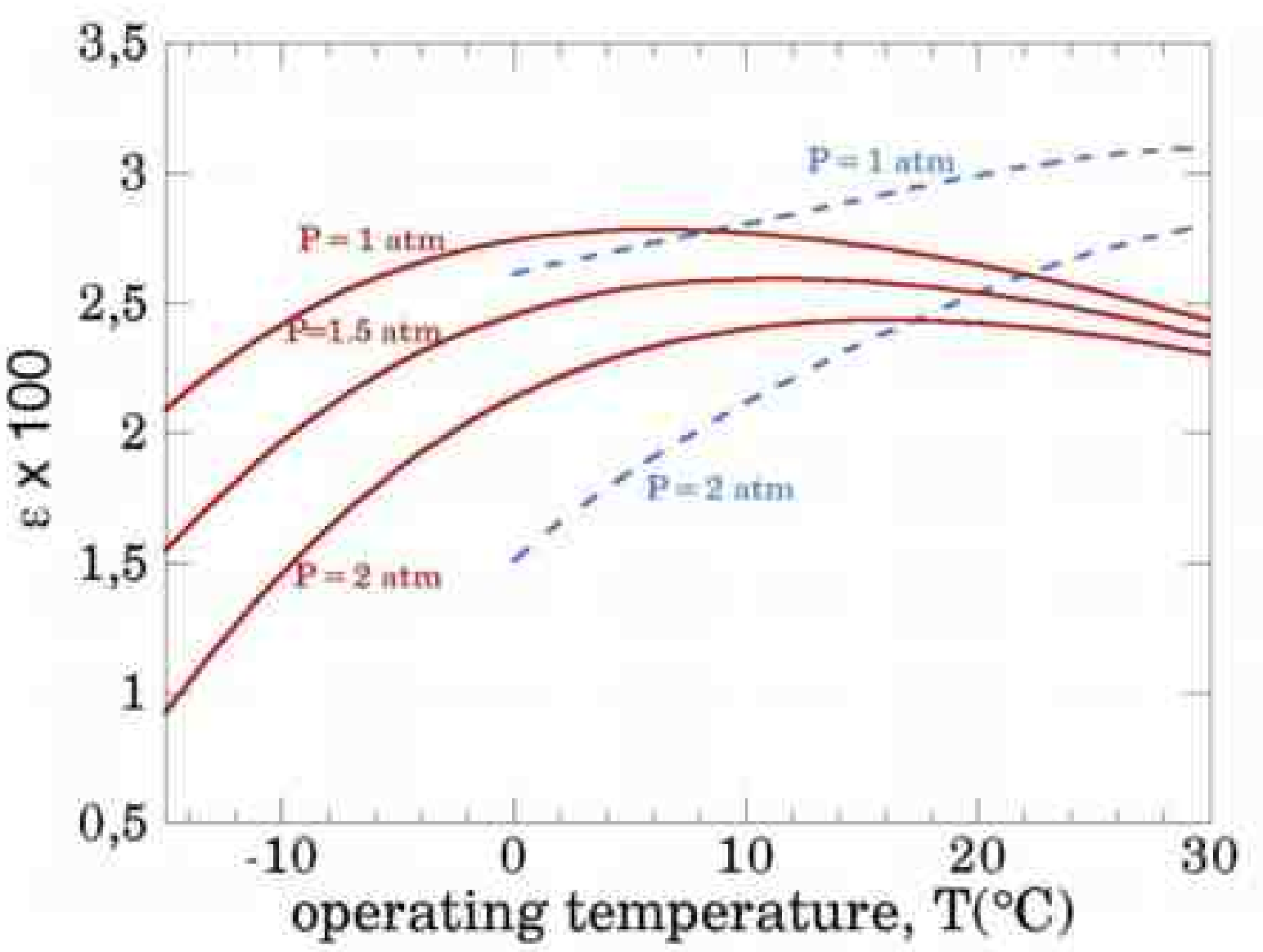}
\caption{Calculated $\epsilon$ for R-115
($C_{2}ClF_{5}$, solid red
lines) and R-12, ($CCl_{2}F_{2}$, dashed blue
lines) as a function
of $P,T$.}
\end{figure}

\newpage
\begin{figure}[tbp]
\epsfxsize = \hsize
\epsfbox {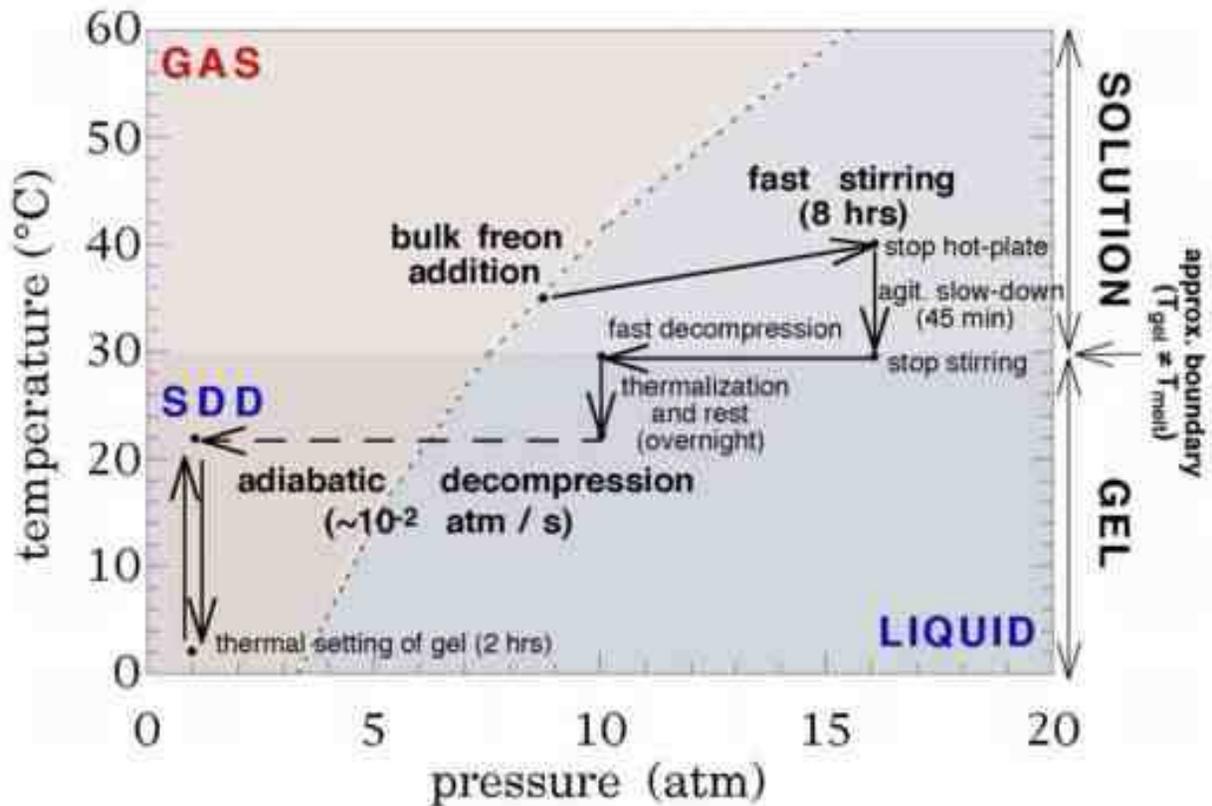}
\caption{Schematic representation of the fabrication
process of a
R-12 SDD. The blue region denotes the liquid phase of the
refrigerant,
pink is for gas. Their boundary (dotted line) traces
$T(P_{V})$. As a result of the adiabatic crossing of $T(P_{V})$,
the refrigerant 
remains in the (superheated) liquid state.
The horizontal division between the sol and gel
states of the
matrix is only approximate, since melting and gelation temperature
are not
the same. The fabrication path changes radically for
other
refrigerants.}
\end{figure}

\newpage
\begin{figure}[tbp]
\epsfxsize = \hsize 
\epsfbox{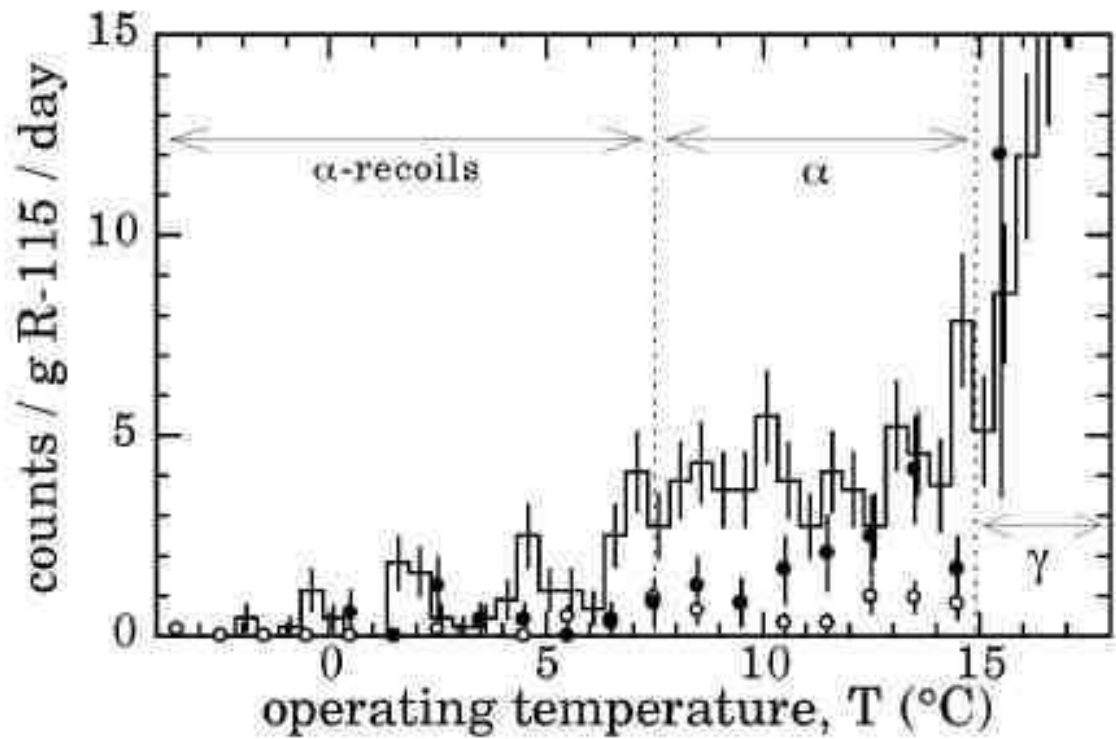}
\caption{SDD background at 90 m.w.e. and
$P\!\!=$2 atm,
following cumulative steps of cleansing; histogram: double
distillation of
water and microfiltration, $\bullet$: single distillation
of refrigerant and glycerin purification, $\circ$: gelatine and
PVP purification. Vertical lines separate three different regimes of
background dominance (see text).}
\end{figure}

\newpage
\begin{figure}[tbp]
\epsfxsize = \hsize
\epsfbox{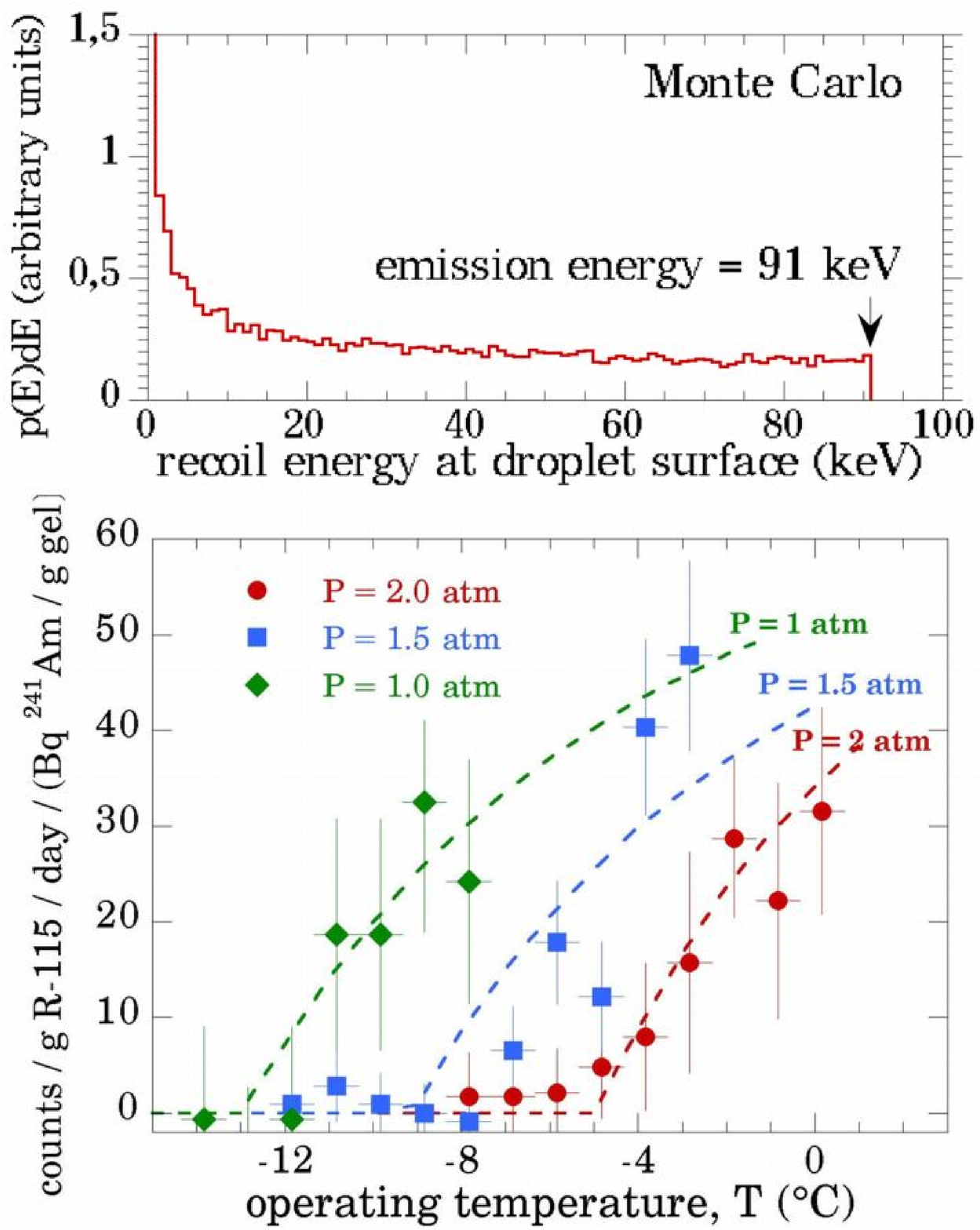}
\caption{High-$dE/dx$ recoiling Np daughters (emitted
simultaneously
to $^{241}$Am alpha
decay in the gel) suffer energy
losses prior to their arrival to a
neighboring droplet, resulting in a 
moderated energy spectrum {(\it top)} that
falls in the same regime where WIMP
recoils are expected (few tens of keV). 
These recoils provide an opportunity to
check the
calculation of the minimum energy required for bubble
formation ($E_{c}$,
Eqs. 1-3): experimentally ({\it bottom}),
the operating temperature
$T_{\alpha
r}$ above which they are able to produce nucleations is seen to
vary
with  $P$
following closely the theoretical
predictions (dashed
lines). }
\end{figure}

\newpage
\begin{figure}[tbp]
\epsfxsize = \hsize
\epsfbox{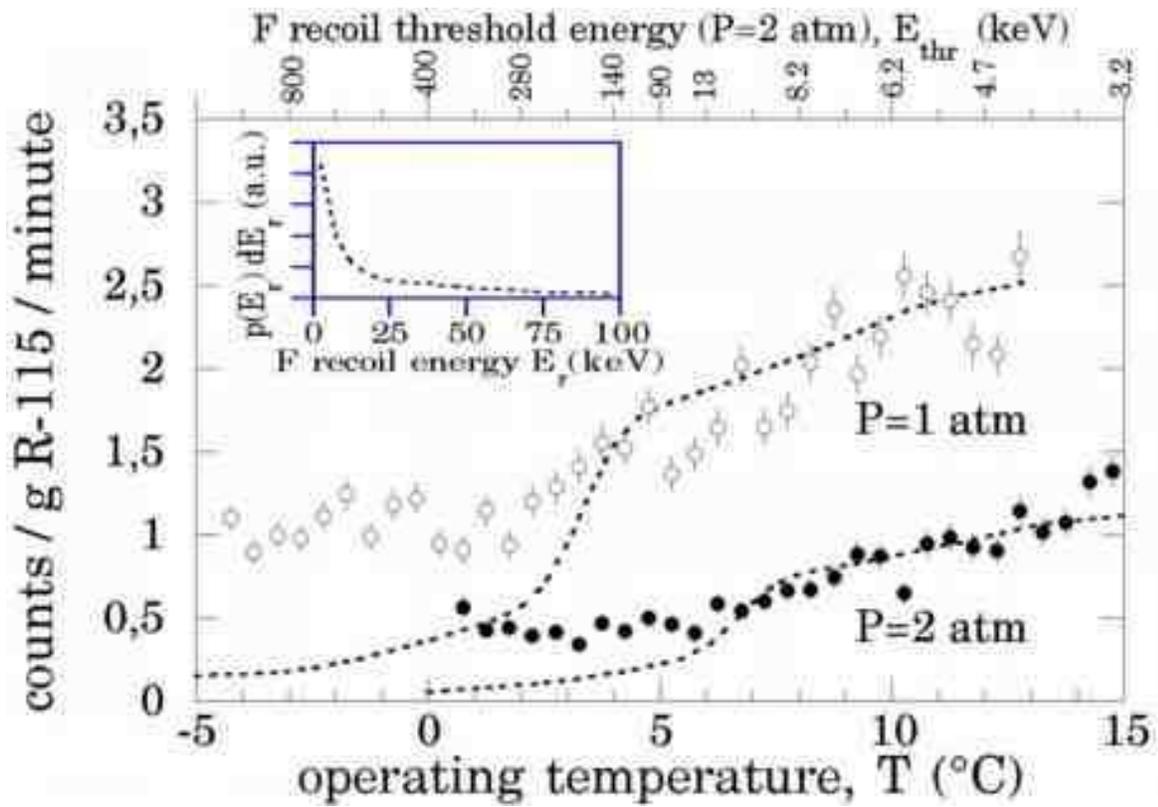}
\caption{$^{252}$Cf neutron calibration of SIMPLE
modules at the
TIS/RP bench (CERN),
compared with Monte Carlo
expectations (dotted lines, see text).
The signal-to-noise ratio
was
$>30$ at all times. {\it Insert}: calculated energy spectrum of F
recoils
during the irradiations.}

\end{figure}

\newpage
\begin{figure}[tbp]
\epsfxsize = \hsize
\epsfbox{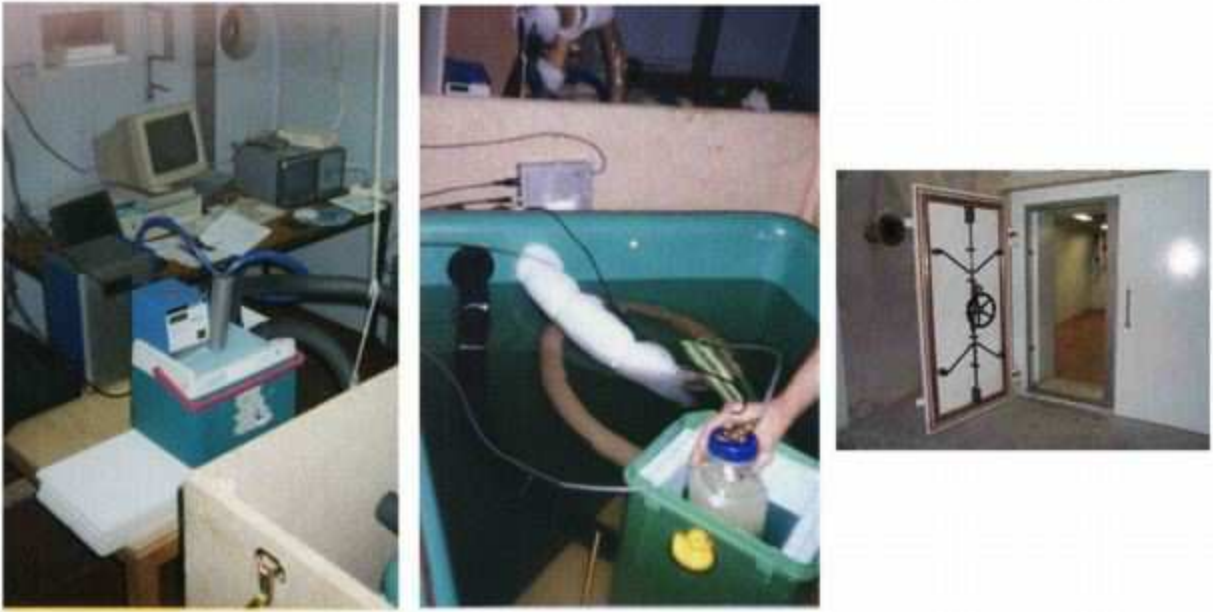}
\caption{SIMPLE at 1,500 m.w.e.. {\it Left:} DAQ system
and water
temperature controlling system. {\it Middle:} first 9.2 g R-115
module
being immersed in the 700 l of water used as neutron moderator;
sound and $T$
insulating layers of the shielding are apparent. {\it
Right:} entrance to the
experimental hall. The Cu-Be contacts that close
the Faraday cage
are visible on the rim of the door.}
\end{figure}

\newpage
\begin{figure}[tbp]
\epsfxsize = \hsize
\epsfbox{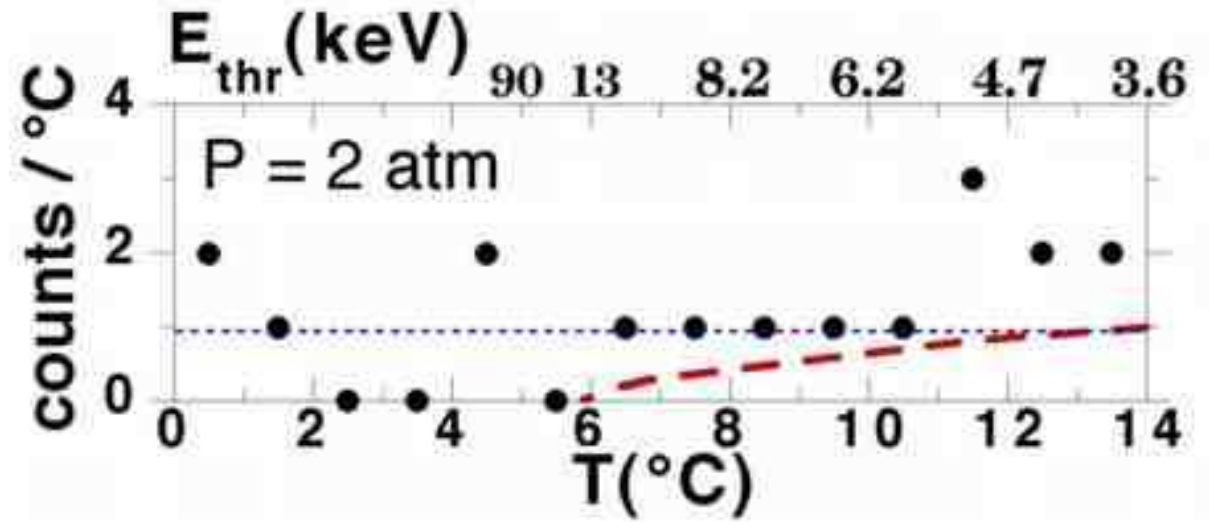}
\caption{Counting rate in the first SIMPLE module
installed in
Rustrel ($9.2\pm0.1$ g R-115,
$\Delta
T=-0.75^{\circ}$C/day). The top axis displays the calculated
threshold
energy for bubble nucleation from
fluorine recoils. The blue dotted line
indicates the average
level of spurious background observed in
refrigerant-free runs (see
text). The red dashed line is the expected
signal (accounting for 34\% sound detection efficiency and F fraction)
from a WIMP of mass $m_{\chi}=$10 GeV and $\sigma_{Wp}=$ 5 pb, i.e.,
at
the limit of sensitivity of the DAMA experiment:
present SIMPLE limits are
largely limited by low statistics.}
\end{figure}

\newpage
\begin{figure}[tbp]
\epsfxsize = \hsize
\epsfbox{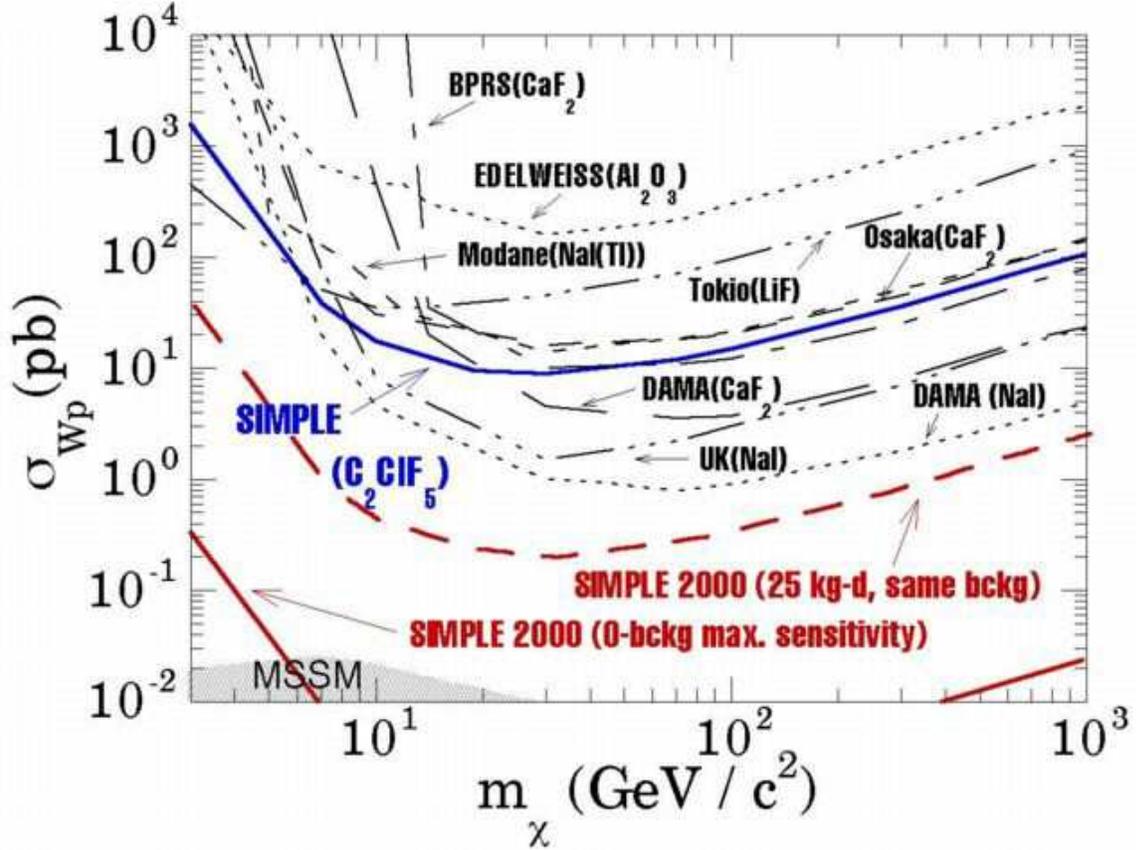}
\caption{95\% C.L. limits on $\sigma_{Wp}$ extracted
from only 0.19
kg-day of SDD exposure, compared with other experiments
[26].
The red lines indicate the expected sensitivity of SIMPLE 2000 after
an
exposure of 25 kg-day, if no improvement in the background is
obtained
(dashed line) or at the maximum reachable level for this exposure 
(zero
background, solid
red line off the scale).
``MSSM'' marks the tip of the region where a
lightest supersymmetric
partner is expected. }
\end{figure}

\newpage
\begin{figure}[tbp]
\epsfxsize = \hsize
\epsfbox{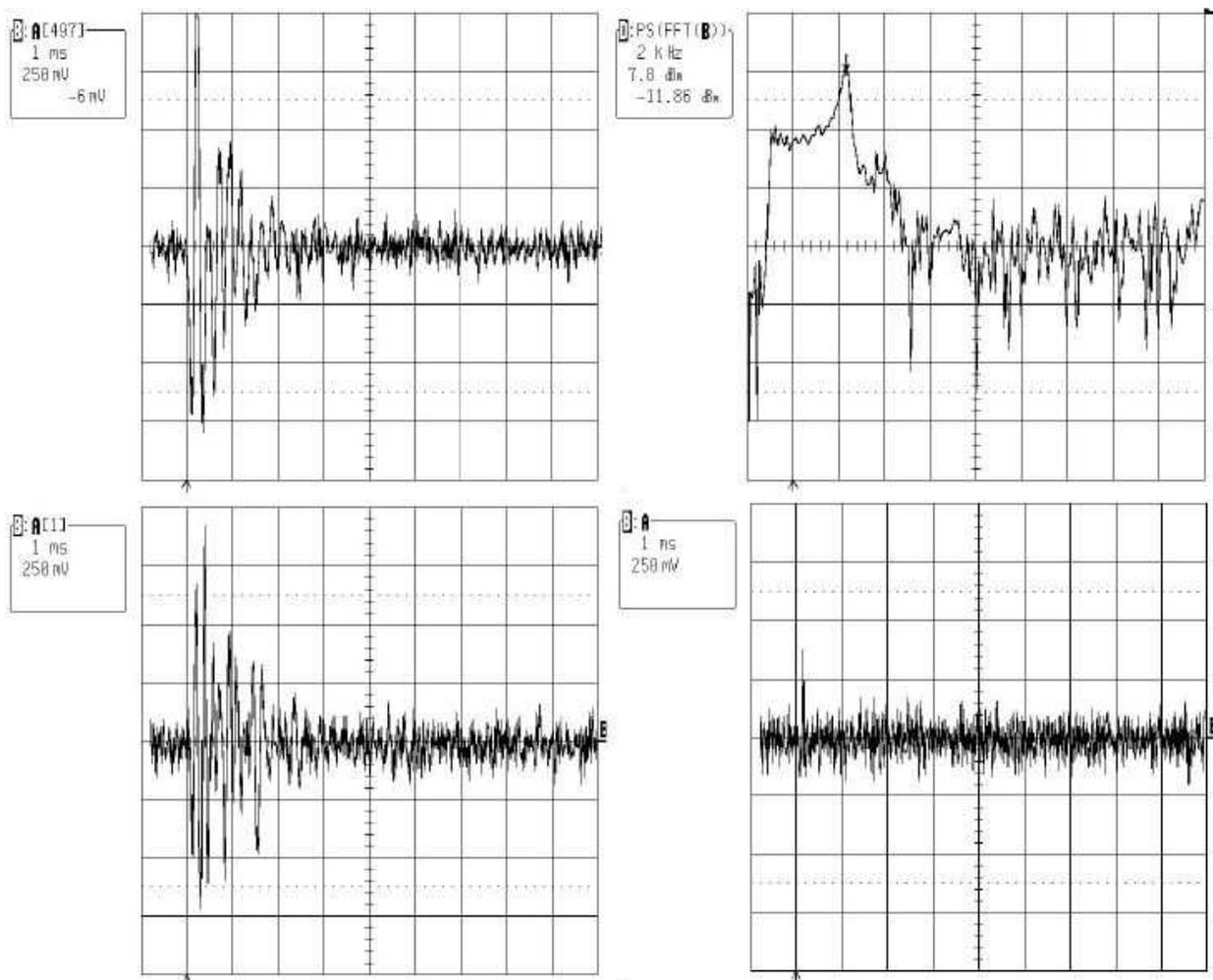}
\caption{Signal and noise in present SIMPLE modules: the
pulse shape
(top-left) corresponds to a typical bubble nucleation, with a
dominant frequency of $\sim 5$ kHz (top-right shows its
Fourier
transform) and time span of few ms. During runs with
refrigerant-free
``dummy'' modules, similar signals (bottom-left) are
observed arising
from pressure microleaks in plastic SDD caps, at a rate of
$\sim 1$/day.
Even at atmospheric pressure, a residual
rate of $\sim
0.3$/day characteristic EM noise events (bottom-right) is
present.
As a first measure, sturdier metallic SDD caps have been built
and all non-essential
equipment (PC, water chiller) is to be moved outside
of the Faraday cage. The sharply-resonant piezoelectric sensors 
presently employed will eventually be substituted by others with a 
flatter spectral response, allowing for univocal identification of 
the nucleation sounds.}
\end{figure}

\newpage
\begin{figure}[tbp]
\epsfxsize = \hsize
\epsfbox{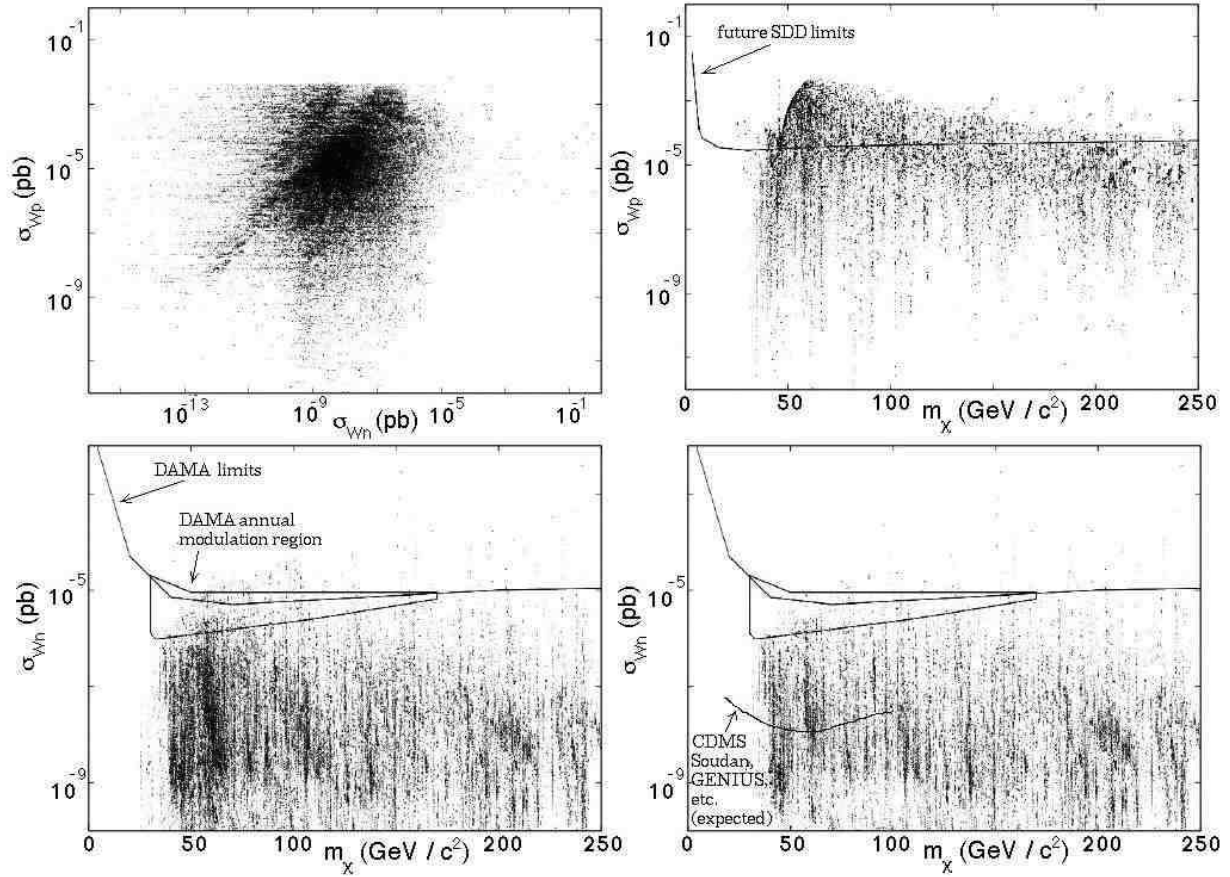}
\caption{A fluorine-rich SDD of modest active mass (O(1) kg), the 
ultimate goal of the SIMPLE program, will be 
sensitive to neutralino WIMP candidates beyond the 
reach of the most ambitious planned cryogenic experiments (see text).}
\end{figure}

\end{document}